\documentclass[11pt]{article}

\usepackage{times}
\usepackage{latexsym}

\usepackage[T1]{fontenc}

\usepackage[utf8]{inputenc}

\usepackage{microtype}

\usepackage{inconsolata}

\usepackage{graphicx}

\usepackage{amsmath}
\usepackage{graphicx}
\usepackage{multicol}
\usepackage{multirow}
\usepackage{booktabs}
\usepackage{amssymb}

\usepackage[final]{acl}
\usepackage{cleveref}
\newif\ifcameraready
\camerareadytrue
\usepackage{graphicx}
\usepackage{bxcoloremoji}

 \usepackage{microtype}
\usepackage{tikz}
\usetikzlibrary{positioning,fit,arrows.meta,calc}

\usepackage[english,bidi=default]{babel} 
\babelfont{rm}{TeXGyreTermesX} 

\title{DialogueSidon: Recovering Full-Duplex Dialogue Tracks\\ from In-the-Wild Dialogue Audio}

\author{
 \textbf{Wataru Nakata\textsuperscript{1,2}},
 \textbf{Yuki Saito\textsuperscript{1,2}},\\
 \textbf{Kazuki Yamauchi\textsuperscript{1}},
 \textbf{Emiru Tsunoo\textsuperscript{1}},
 \textbf{Hiroshi Saruwatari\textsuperscript{1}},
\\
\\
 \textsuperscript{1}The University of Tokyo, Tokyo, Japan, \\
 \textsuperscript{2}National Institute of Advanced Industrial Science and Technology (AIST), Tokyo, Japan
\\
 \small{
   \textbf{Correspondence:} \href{mailto:nakata-wataru855@g.ecc.u-tokyo.ac.jp}{nakata-wataru855@g.ecc.u-tokyo.ac.jp}
 }
}

\begin{document}

\maketitle
\begin{abstract}
Full-duplex dialogue audio, in which each speaker is recorded on a separate track, is an important resource for spoken dialogue research, but is difficult to collect at scale. Most in-the-wild two-speaker dialogue is available only as degraded monaural mixtures, making it unsuitable for systems requiring clean speaker-wise signals. We propose DialogueSidon, a model for joint restoration and separation of degraded monaural two-speaker dialogue audio. DialogueSidon combines a variational autoencoder (VAE) operates on the speech self-supervised learning (SSL) model feature, which compresses SSL model features into a compact latent space, with a diffusion-based latent predictor that recovers speaker-wise latent representations from the degraded mixture. Experiments on English, multilingual, and in-the-wild dialogue datasets show that DialogueSidon substantially improves intelligibility and separation quality over a baseline, while also achieving much faster inference.
\end{abstract}
\section{Introduction}

Building natural spoken dialogue systems remains a key challenge in dialogue research. Recent systems have begun to exhibit human-like conversational behaviors such as backchannels, overlap, and flexible turn-taking, enabling fluent human-AI conversation~\cite{nguyen-etal-2023-generative,defossez2024moshispeechtextfoundationmodel, roy2026personaplexvoicerolecontrol}. However, modeling such phenomena robustly still requires large amounts of suitable conversational speech data.
\begin{figure}[t]
    \centering
    \includegraphics[width=0.85\linewidth]{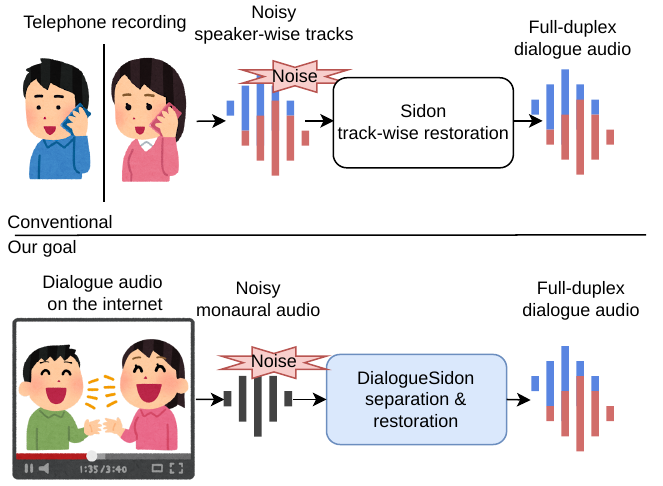}
    \vspace{-3mm}
    \caption{Comparison of current full-duplex dialogue audio acquisition and our goal}
    \label{fig:goal}
    \vspace{-5mm}
\end{figure}

In particular, full-duplex dialogue recordings are important resources for spoken dialogue research. In these recordings, two speakers are recorded on separate tracks, making it possible to analyze and model conversational behaviors while preserving clean speaker-wise signals. Such data are especially valuable for studying natural interaction and for developing human-like dialogue systems.

Despite their importance, full-duplex dialogue data are difficult to collect at scale. A common approach is to record telephone conversations~\cite{cieri-etal-2004-fisher,ldc2008_cabank_callhome_english} or controlled two-party interactions~\cite{seamless_interaction,dailytalk}, but such pipelines are costly to build, often suffer from latency and channel artifacts, and are not easily scalable. Another possibility is to synthesize conversations using text-to-speech systems~\cite{defossez2024moshispeechtextfoundationmodel}. However, because these systems are usually driven by turn-based text dialogues, the resulting audio often lacks spontaneous interactional phenomena such as natural overlap, backchannels, and timing variability.

Meanwhile, large-scale in-the-wild audio from the Internet provides a potentially rich source of conversational speech~\cite{nakata2026jchatjapaneselargescalespoken}. Internet recordings contain diverse speakers, speaking styles, and interaction patterns that are difficult to reproduce in controlled data collection. 
In fact, such data are already used for tasks such as large-scale (SSL)~\cite{usm}, automatic speech recognition~\cite{whisper, peng25c_interspeech}, and speech synthesis~\cite{koizumi23_interspeech}.

However, such recordings are difficult to use directly for training full-duplex dialogue systems.
First, Internet audio is heavily degraded. Recordings may contain background music, environmental noise, reverberation, clipping, and compression. Second, Internet dialogue audio is usually available only as mixed recordings rather than isolated speaker-wise tracks. Jointly addressing both problems---restoring signal quality and separating speakers---is therefore necessary to recover speaker-wise signals from in-the-wild dialogue audio.

A natural approach is to apply speech separation~\cite{convtasnet}. However, existing separation methods are typically developed for mixtures of monologue clean speech and often do not generalize well to spontaneous conversational speech, where overlap patterns, prosody, and interaction timing are substantially different. 

Speech restoration methods~\cite{liu22y_interspeech,karita2025miipher,nakata2026sidonfastrobustopensource}, on the other hand, have recently shown strong performance in removing noise and recording artifacts. In fact, some of open corpora are based on the restored samples by speech restoration models~\cite{koizumi23_interspeech,ma24c_interspeech,emilia}. 
Inspired by these approaches, the joint modeling of speech restoration and separation has been proposed~\cite{asai2026genesesunifiedgenerativespeech,noisySS}.
Nevertheless, existing work has mainly focused on monologue and monaural speech rather than mixed dialogue recordings.
This limits their applicability to full-duplex dialogue recovery from in-the-wild two-speaker dialogue audio.

In this paper, we propose \textit{DialogueSidon}, a model for joint restoration and separation of in-the-wild dialogue as shown in \Cref{fig:goal}. The key idea is to extend the previous open-source speech restoration model Sidon~\cite{nakata2026sidonfastrobustopensource} to the dialogue setting by incorporating speaker separation, allowing degraded mixed conversational recordings to be transformed into cleaner speaker-wise dialogue signals by leveraging SSL model pretrained on large-scale speech dataset. Experimental results show that DialogueSidon improves both restoration and separation quality compared with conventional unified restoration and separation baselines, while also achieving faster inference.

Our contributions are as follows:
(i) We propose DialogueSidon, a model that jointly restores and separates degraded monaural two-speaker dialogue audio into clean speaker-wise tracks.
(ii) DialogueSidon combines an SSL-VAE latent space with a diffusion-based latent predictor, enabling speaker-wise recovery from degraded conversational mixtures.
(iii) We show through experiments on English, multilingual, and in-the-wild dialogue data that DialogueSidon improves content preservation and separation quality over a baseline while providing substantially faster inference.
\ifcameraready
Audio samples\footnote{\url{https://hf.co/spaces/Wataru/dsidonsamples}}, code\footnote{\url{https://github.com/sarulab-speech/Sidon}}, and a live demo\footnote{\url{https://hf.co/spaces/sarulab-speech/DialogueSidon-demo}} are publicly available.
\fi

\begin{figure*}[t]
\centering
\includegraphics[width=0.8\linewidth]{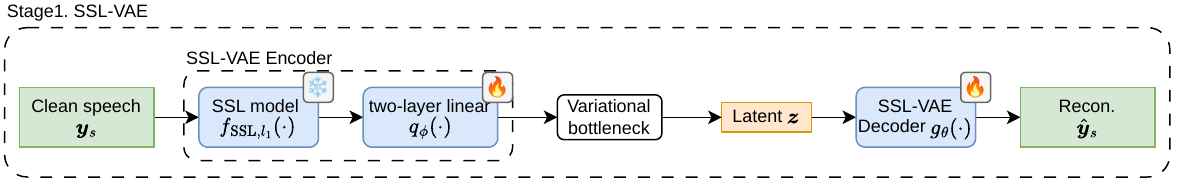}

\includegraphics[width=0.8\linewidth]{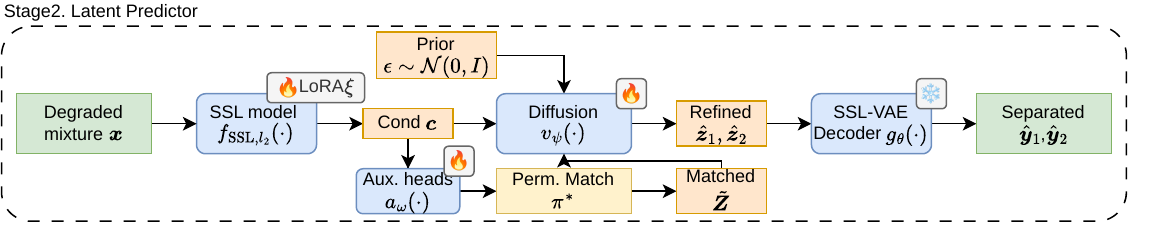}

\caption{Overview of DialogueSidon training.
\coloremoji{❄️} indicates frozen modules and \coloremoji{🔥} indicates trained modules.}
\label{fig:model}
\vspace{-5mm}
\end{figure*}

\section{Related Work}
\subsection{Spoken Dialogue Language Modeling}

Recent spoken dialogue language models~\cite{nguyen-etal-2023-generative,defossez2024moshispeechtextfoundationmodel,roy2026personaplexvoicerolecontrol,veluri-etal-2024-beyond} can capture acoustic and temporal phenomena such as overlap, backchannels, and turn-taking timing by modeling spoken interactions directly. However, these models still lag behind human conversation~\cite{nguyen-etal-2023-generative,defossez2024moshispeechtextfoundationmodel,roy2026personaplexvoicerolecontrol}, partly due to the scarcity of large-scale full-duplex dialogue data---recordings where each speaker is on a separate track. Existing datasets are typically limited to telephone corpora such as Fisher~\cite{cieri-etal-2004-fisher} (2k hours) and proprietary recordings~\cite{defossez2024moshispeechtextfoundationmodel}, far smaller than the scale used in recent speech generation research which frontier models use millions of hours of speech~\cite{usm,peng25c_interspeech}. This data scarcity motivates our work on converting Internet audio into usable full-duplex dialogue data.

\subsection{Speech Separation and Restoration}

Speech separation~\cite{convtasnet,wang23tf-gridnet,floss} aims to recover individual speaker signals from multi-speaker mixtures. However, most methods are developed on clean monologue mixtures and do not generalize well to in-the-wild conversational audio with background music, noise, reverberation, and compression artifacts.

Speech restoration removes degradations from corrupted recordings~\cite{liu22y_interspeech,miipher,karita2025miipher,nakata2026sidonfastrobustopensource}, making it useful for large-scale dataset cleansing~\cite{koizumi23_interspeech,ma24c_interspeech}. Recent methods such as Sidon~\cite{nakata2026sidonfastrobustopensource} perform restoration in the latent space of SSL model, improving robustness to diverse in-the-wild degradations. However, they mainly target single-speaker speech and do not address speaker separation.

A straightforward cascade of restoration and separation is generally inadequate for this setting. Applying restoration first can suppress or distort overlapping speech, because existing restoration models are typically designed for monologue and tend to treat overlap as corruption to be removed. Applying separation first is also unreliable, since most separation models are not designed for heavily degraded in-the-wild mixtures. This motivates unified modeling of separation and restoration rather than a simple cascade.

More recently, several studies have explored unified modeling of speech separation and restoration~\cite{noisySS,asai2026genesesunifiedgenerativespeech}. These approaches are promising because both problems must be addressed to recover usable speaker-wise signals from degraded mixtures. However, existing methods are primarily developed for mixtures of monologue speech and may not generalize well to in-the-wild dialogue audio, where overlap patterns, backchannels, and interaction timing differ substantially from monologue mixtures.


\section{DialogueSidon}
DialogueSidon extends the speech restoration model Sidon~\cite{nakata2026sidonfastrobustopensource} to the dialogue setting by introducing speaker-wise recovery from degraded conversational mixtures. Given a degraded monaural two-speaker dialogue mixture, the goal is to reconstruct a clean waveform for each speaker track. As shown in \Cref{fig:model}, the model consists of two components: (i) an SSL-VAE that defines a compact latent space for clean speech, and (ii) a diffusion-based latent predictor that estimates the speaker-wise latent representations from the degraded mixture. The predicted latents are then decoded into separated clean waveforms.

Formally, given a degraded monaural two-speaker dialogue mixture with $T$ samples, $\boldsymbol{x} \in \mathbb{R}^{T}$, our goal is to recover the corresponding clean full-duplex waveforms $\boldsymbol{y}_1, \boldsymbol{y}_2 \in \mathbb{R}^{T}$, where each waveform corresponds to one speaker. In this work, we focus on the two-speaker case.

DialogueSidon first maps each target signal $\boldsymbol{y}_s$ to a compact latent representation $\boldsymbol{z}_s \in \mathbb{R}^{L \times D}$ using the SSL model followed by two layer linear layers with ReLU activation, where $s \in \{1,2\}$, $L$ is the latent sequence length, and $D$ is the latent dimensionality. The latent predictor is then trained to estimate $\hat{\boldsymbol{z}}_1, \hat{\boldsymbol{z}}_2$ from the degraded mixture $\boldsymbol{x}$. Finally, the SSL-VAE decoder reconstructs the corresponding waveforms $\hat{\boldsymbol{y}}_1, \hat{\boldsymbol{y}}_2$ from the predicted latents.

Training is performed in two stages. In the first stage, the SSL-VAE is trained to construct a compact and speech-relevant latent space from clean dialogue tracks. In the second stage, the latent predictor is trained in this latent space to recover the speaker-wise clean representations from degraded mixed dialogue audio. At inference time, the latent predictor first estimates the latent representation for each speaker track from the degraded input, and the SSL-VAE decoder then reconstructs the final waveforms.

\subsection{SSL-VAE}\label{sec:sslvae}

The SSL-VAE is designed to compress SSL model representations into a compact latent space suitable for diffusion modeling. Previous studies on speech restoration~\cite{miipher,karita2025miipher,Guimares2025DiTSEHG} have shown that features extracted from large SSL models are effective for representing speech content and acoustic characteristics. However, directly applying diffusion models to such high-dimensional (e.g., 1,536-dimensional) feature spaces is computationally expensive. To address this issue, we adopt a latent diffusion framework~\cite{hiresldm} in which SSL features are first compressed with a pretrained VAE encoder.

Let $f_{\text{SSL},l_1}(\cdot)$ denote a frozen SSL $l_1$-th layer hidden feature. Given a clean monaural waveform $\boldsymbol{y}$, we first compute a $D_h$-dimensional intermediate hidden representation $\boldsymbol{h} = f_{\text{SSL},l_1}(\boldsymbol{y})$, where $\boldsymbol{h} \in \mathbb{R}^{L \times D_h}$.
The trainable encoder $q_{\phi}$ then applies a two-layer linear layers with ReLU activation followed by a variational bottleneck to obtain a compact latent representation $\boldsymbol{z} \sim q_{\phi}(\boldsymbol{z} \mid \boldsymbol{h})$ with dimension $D < D_h$. The latent variable $\boldsymbol{z} \in \mathbb{R}^{L \times D}$ serves as the target representation for the second-stage latent predictor.

The trainable decoder $g_{\boldsymbol{\theta}}(\cdot)$ reconstructs the original waveform from the latent representation: $\hat{\boldsymbol{y}} = g_{\boldsymbol{\theta}}(\boldsymbol{z})$. The training objective consists of a reconstruction term, adversarial loss, together with the standard Kullback--Leibler regularization similar to previous work~\cite{shi2025samaudiosegmentaudio}: $\mathcal{L}_{\mathrm{VAE}} = \mathcal{L}_{\mathrm{rec}}(\boldsymbol{y}, \hat{\boldsymbol{y}}) + \mathcal{L}_{\mathrm{adv}}(\boldsymbol{y}, \hat{\boldsymbol{y}}) + \beta \, D_{\mathrm{KL}}(q_{\phi}(\boldsymbol{z} \mid \boldsymbol{h})\,\|\,p(\boldsymbol{z}))$, where $p(\boldsymbol{z})$ is a prior distribution and $\beta$ controls the strength of regularization.

Following prior work~\cite{nakata2026sidonfastrobustopensource}, we use $l_1=8$-th layer hidden feature of  w2v-BERT 2.0~\cite{Communication2023SeamlessME} as the SSL feature extractor. This model is trained on 4.5 million hours of speech covering 143 languages and has been shown to be effective for speech-related tasks including speech translation~\cite{Communication2023SeamlessME} and speech restoration~\cite{nakata2026sidonfastrobustopensource}. For waveform reconstruction and discriminator used for adversarial training, we use the Descript Audio Codec~\cite{kumar2023highfidelity} decoder and discriminator, which is based on a HiFi-GAN-style vocoder~\cite{Kong2020HiFi-GAN:Synthesis} with Snake activation~\cite{snake}.

\subsection{Latent Predictor}

The role of the latent predictor is to estimate clean speaker-wise latent representations from a degraded monaural dialogue mixture.  The latent predictor takes $\boldsymbol{x}$ as input and predicts $\hat{\boldsymbol{z}}_1, \hat{\boldsymbol{z}}_2$, which are then decoded by the SSL-VAE decoder to obtain the final reconstructed waveforms.

A straightforward approach would be to directly regress the target latents from the degraded input. However, in our preliminary experiments, such deterministic objectives tended to oversmooth the latent trajectories and often failed to preserve spoken content and overlapping speech. To better model the uncertainty inherent in degraded conversational mixtures, we instead adopt a diffusion-based latent predictor.
This approach is reported to be effective in other speech processing tasks~\cite{shi2025samaudiosegmentaudio,Guimares2025DiTSEHG}.

\paragraph{Conditioning representation}
We first extract a conditioning representation $\boldsymbol{c} = f_{\text{SSL},l_2,\xi}(\boldsymbol{x}) \in \mathbb{R}^{L \times D_h}$ from the degraded input mixture, where $f_{\mathrm{SSL},l_2, \xi}(\cdot)$ is an $l_2$-th layer SSL model feature fine-tuned with low-rank adapter  (LoRA)~\cite{hu2022lora} parameterized by $\xi$.
This representation captures the linguistic and acoustic information in the degraded mixture and is used to condition both the auxiliary latent prediction and the diffusion model.

For conditioning, we use the $l_2=13$-th layer hidden feature of w2v-BERT 2.0. The LoRA adapter is applied to the last linear layer of the feed-forward network in each Conformer block, adapting the model to degraded mixtures while preserving pretrained knowledge.

\paragraph{Auxiliary latent prediction}
A key challenge in speaker-wise latent prediction is permutation ambiguity: the two output tracks are unordered, therefore the model must not only predict the correct latent content, but also resolve speaker assignment. To reduce the burden on the diffusion model, we introduce auxiliary heads parametrized by $\omega$ and denoted $a_{\omega}$ that produce coarse speaker-wise latent estimates $\tilde{\boldsymbol{z}}_1, \tilde{\boldsymbol{z}}_2 = a_{\omega}(\boldsymbol{c})$ directly from the conditioning representation.
These auxiliary heads are trained with permutation-invariant training~\cite{pit} to encourage them to produce speaker-wise estimates that are close to the target latents, while allowing for flexible speaker assignment. By providing these coarse estimates, we guide the diffusion model towards better predictions and help it resolve permutation ambiguity.
These auxiliary predictions are matched against the ground-truth latents in a permutation-invariant manner~\cite{pit}: $\pi^\star = \arg\min_{\pi \in \mathcal{S}_2} \sum_{s=1}^{2} \| \tilde{\boldsymbol{z}}_s - \boldsymbol{z}_{\pi(s)} \|_1$, where $\mathcal{S}_2$ is the set of all permutations over two speakers. The resulting permutation $\pi^\star$ defines the alignment between predicted and target speaker tracks.
Using this alignment, the auxiliary latent prediction loss is $\mathcal{L}_{\mathrm{aux}} = \sum_{s=1}^{2} \| \tilde{\boldsymbol{z}}_s - \boldsymbol{z}_{\pi^\star(s)} \|_1$. The same permutation $\pi^\star$ is then used to define the speaker ordering for the diffusion model, preventing it from suffering from the permutation ambiguity.
\paragraph{Diffusion-based latent refinement}
After permutation matching, we jointly model the two speaker latents by stacking them into $\boldsymbol{Z}= \mathrm{stack}(\boldsymbol{z}_{\pi^\star(1)}, \boldsymbol{z}_{\pi^\star(2)}) \in \mathbb{R}^{L \times 2D} $ and similarly $\tilde{\boldsymbol{Z}} = \mathrm{stack}(\tilde{\boldsymbol{z}_1}, \tilde{\boldsymbol{z}_2})\in \mathbb{R}^{L \times 2D}$ for the auxiliary predictions. The stacking is performed along the latent dimension $D$. This joint representation allows the diffusion model to capture inter-speaker dependencies.

We use the v-prediction~\cite{salimans2022progressive} reparametrization of diffusion model training following previous work~\cite{Guimares2025DiTSEHG}. The forward diffusion process is $\boldsymbol{Z}_t = \alpha_t \boldsymbol{Z} + \sigma_t \boldsymbol{\epsilon}$ with $\boldsymbol{\epsilon} \sim \mathcal{N}(0, I)$. Here, $t \sim \mathcal{U}(0, 1)$ represents the continuous diffusion time step, while the scalar functions $\alpha_t$ and $\sigma_t \in (0,1)$  denote the signal and noise schedules, respectively. The v-prediction target is defined as $\boldsymbol{v}_t = \alpha_t \boldsymbol{\epsilon} - \sigma_t \boldsymbol{Z}$. The diffusion model $v_{\boldsymbol{\psi}}(\cdot)$ parameterized by $\boldsymbol{\psi}$, conditioned on the SSL feature $\boldsymbol{c}$ and auxiliary predictions $\tilde{\boldsymbol{Z}}$, predicts this velocity: $\hat{\boldsymbol{v}}_t = v_{\boldsymbol{\psi}}(\boldsymbol{Z}_t, t, \boldsymbol{c}, \tilde{\boldsymbol{Z}})$.
The diffusion loss is $\mathcal{L}_{\mathrm{diff}} = \mathbb{E}_{\boldsymbol{Z}, \boldsymbol{\epsilon}, t} [ \| \boldsymbol{v}_t - v_{\psi}(\boldsymbol{Z}_t, t, \boldsymbol{c}, \tilde{\boldsymbol{Z}}) \|_2^2 ]$.
This decomposition separates speaker alignment (auxiliary heads) from latent refinement (diffusion), allowing each component to focus on a distinct sub-problem.

\paragraph{Training objective}
The latent predictor is trained using the sum of the auxiliary latent loss and the diffusion loss: $\mathcal{L}_{\mathrm{latent}} = \mathcal{L}_{\mathrm{aux}} + \lambda_{\mathrm{diff}} \mathcal{L}_{\mathrm{diff}}$, where $\lambda_{\mathrm{diff}}$ controls the weight of the second term.

\paragraph{Inference}
At inference time, the auxiliary heads produce coarse speaker-wise estimates from $\boldsymbol{c}$, and the diffusion model refines them via the reverse process. The SSL-VAE decoder then reconstructs clean waveforms $\hat{\boldsymbol{y}}_s = g_{\theta}(\hat{\boldsymbol{z}}_s)$ for $s \in \{1,2\}$.

 \begin{table}[t]
  \centering
  \caption{Durations of dataset used for training.}
  \vspace{-2mm}
  \footnotesize
  \label{tab:dataset_stats}
  \scalebox{0.9}{
  \begin{tabular}{lr}
  \toprule
  \textbf{Dataset} &  \textbf{Duration (h)} \\
  \midrule
  CALLHOME German~\cite{ldc2008_cabank_callhome_english}   &    58.1 \\
  CALLHOME English~\cite{ldc2008_cabank_callhome_english}  &    76.9 \\
  CALLHOME Japanese~\cite{ldc2008_cabank_callhome_japanese} &    49.3 \\
  CALLHOME Spanish~\cite{ldc2008_cabank_callhome_spanish}  &    43.6 \\
  CALLHOME Mandarin~\cite{ldc2008_cabank_callhome_chinese} &    39.5 \\
  Fisher~\cite{cieri-etal-2004-fisher}            & 1,958.5 \\
  \midrule
  \textbf{Total}    & 2,225.9 \\
  \bottomrule
  \end{tabular}
  }
  \vspace{-5mm}
\end{table}
\section{Experiments}
We evaluate the speech restoration and separation capabilities of DialogueSidon in English, multilingual, and in-the-wild settings.
All models are trained on Sidon-restored telephone conversational corpora, which provide clean speaker-wise tracks necessary for supervised training but are not in-the-wild data.
The Sidon-restored samples (48~kHz) were downsampled to 24~kHz.
Evaluation is conducted on three corpora: Switchboard (SWB)~\cite{godfrey1993_switchboard1_release2} for in-domain English telephone audio, CallFriend~\cite{canavan1996_callfriend_german,canavan1996_callfriend_canadian_french,canavan1996_callfriend_japanese,canavan1996_callfriend_spanish_non_caribbean,tracey2025_bolt_callfriend_callhome_mandarin} for multilingual telephone audio, and OpenDialog~\cite{zhu2025zipvoicedialognonautoregressivespokendialogue} for in-the-wild Internet audio, to assess both in-domain and out-of-domain generalization.

\subsection{Experimental conditions}
The statistics of the datasets used for the experiment are shown in \Cref{tab:dataset_stats}.
We used a combination of the Fisher English dialogue corpus and the English, German, Japanese, Spanish, and Mandarin variants of the CALLHOME corpus. These are conversational corpora consisting of telephone conversations on various topics.
For preprocessing, the recordings were restored with Sidon~\cite{nakata2026sidonfastrobustopensource}, as they are telephone recordings and thus contain a wide range of noise with limited audio fidelity.
\paragraph{Data preparation}
To train a generalized speech restoration and separation model, covering a wide range of degradations is important, as we cannot make assumptions about the specific degradations present in in-the-wild dialogue audio. Therefore, similar to previous work~\cite{saijo25_interspeech}, we used diverse sets of degradations.
Each degradation was applied independently to each track with the probability of 0.5 before mixing into a monaural dialogue in the final step.
The details of the degradations are reported in \Cref{sec:deg}.

The noising pipeline was applied four times to each dialogue session with different random seeds. As a result, we obtained approximately 8{,}902 hours of paired (clean, degraded) dialogue data.

\paragraph{DNN architecture}
For the SSL feature extractor, we use w2v-BERT 2.0 as described in \cref{sec:sslvae}. We investigate the effect of the latent channel dimensionality in SSL-VAE by varying $D \in \{8, 16, 32, 64, 128\}$ while keeping the temporal sequence length fixed. All models produce 24\,kHz output audio.

For the latent predictor, the auxiliary heads consist of two separate linear projection layers applied to the conditioning feature $\boldsymbol{c}$, one per speaker track. In the SSL model used in the latent predictor, we performed fine-tuning with a LoRA adapter with rank $r=64$ and scaling factor $\alpha=16$. The architecture of diffusion model is a Diffusion Transformer (DiT)~\cite{Peebles_2023_ICCV} with hidden dimension 768, eight layers, 12 attention heads, intermediate size 3,072. Rotary positional encoding~\cite{rope} is used in the DiT. The conditioning feature $\boldsymbol{c}$ and the stacked auxiliary latent prediction $\tilde{\boldsymbol{Z}}$ and noised input $\boldsymbol{Z}_t$ are concatenated along the latent dimension and provided. We use a linear noise schedule with 1,000 diffusion steps during training.
At inference time, we used DPM-Solver++~\cite{Lu_2025} with 30 steps for sampling.

\paragraph{Training details}
The SSL-VAE is trained for two days with a batch size of 32 using the AdamW\cite{loshchilov2018decoupled} optimizer with a learning rate of $1 \times 10^{-4}$. The exponential decay learning rate schedule is used where the learning rate is multiplied by a factor $\gamma=0.999996$ every step. The KL regularization weight is set to $\beta = 1\times 10^{-5}$. The latent predictor is trained for two days with a batch size of 64, a learning rate of $1 \times 10^{-4}$ with 2,000 warm-up steps, and diffusion loss weight $\lambda_{\mathrm{diff}} = 1.0$. All experiments are conducted on eight NVIDIA H100 GPUs.

\begin{table*}[!ht]
\caption{
Comparison of latent size on the SWB evaluation set.\label{tab:latent_size}
We compare DialogueSidon with different latent dimensionalities $D$ against Noisy, Sidon, and GENESES.
$^\dagger$Noisy and Sidon do not perform speech separation.
\textbf{Bold} and \underline{underline} denote the best and worst among separation methods, respectively.
}
\vspace{-3mm}
\centering
\footnotesize
\scalebox{0.9}{
\begin{tabular}{l|c|cc|c|c}
\toprule
Method & WER (\%) $\downarrow$ & NISQA $\uparrow$ & DNSMOS $\uparrow$ & Spk.\ Sim.\ $\uparrow$ & VAD Acc.\ $\uparrow$ \\
& \textit{\scriptsize Content} & \multicolumn{2}{c|}{\textit{\scriptsize Recording Quality}} & \textit{\scriptsize Speaker} & \textit{\scriptsize Separation} \\
\midrule
Noisy$^\dagger$ & 60.810 & 2.315 & 3.005 & --- & --- \\
Sidon$^\dagger$ & 57.470 & 3.942 & 3.999 & 0.934 & 0.730 \\
\midrule
GENESES (orig) & \underline{79.990} & 3.507 & \underline{3.582} & \underline{0.803} & \underline{0.812} \\
GENESES & 33.540 & \textbf{3.720} & \textbf{3.680} & 0.853 & 0.923 \\
DialogueSidon ($D=8$) & 16.550 & 3.394 & 3.586 & 0.887 & 0.936 \\
DialogueSidon ($D=16$) & 14.950 & \underline{3.370} & 3.653 & \textbf{0.888} & \textbf{0.939} \\
DialogueSidon ($D=32$) & \textbf{14.390} & 3.453 & 3.641 & 0.887 & 0.938 \\
DialogueSidon ($D=64$) & 15.040 & 3.443 & 3.642 & 0.887 & \textbf{0.939} \\
DialogueSidon ($D=128$) & 22.720 & 3.393 & 3.630 & 0.876 & 0.906 \\
\bottomrule
\end{tabular}
}
\vspace{-5mm}
\end{table*}

\begin{table}[!ht]
\label{tab:mos}
\caption{
MOS test results on the SWB evaluation set with their standard deviation.
All pairwise differences are statistically significant ($p < 0.05$, two-sided $t$-test).
}
\vspace{-3mm}
\centering
\footnotesize
\scalebox{0.9}{
\begin{tabular}{lc}
\toprule
Method & MOS $\uparrow$ \\
\midrule
Noisy & 2.815 $\pm$ 0.999 \\
Sidon & 3.289 $\pm$ 1.167 \\
GENESES & 3.482 $\pm$ 1.195 \\
DialogueSidon & \textbf{3.895 $\pm$ 0.948} \\
\bottomrule
\end{tabular}
}
\vspace{-5mm}
\end{table}

\paragraph{Evaluation data}

\textit{SWB.} We use the SWB corpus, a collection of telephone conversations in English. We randomly select 100 dialogue sessions, each cropped to 20 seconds for evaluation. Since SWB and the Fisher training corpus are both telephone-band conversational corpora, SWB results reflect in-domain performance.

\textit{CallFriend.} For multilingual evaluation, we use the multilingual variant of CallFriend corpus~\cite{canavan1996_callfriend_german,canavan1996_callfriend_canadian_french,canavan1996_callfriend_japanese,canavan1996_callfriend_spanish_non_caribbean,tracey2025_bolt_callfriend_callhome_mandarin}, which includes telephone conversations in German, French (Quebec), Japanese, Spanish, and Mandarin. We randomly select 20 dialogue sessions, each cropped to 20 seconds from each language variety for evaluation.

\textit{OpenDialog.} For in-the-wild evaluation, we use the English train set of OpenDialog~\cite{zhu2025zipvoicedialognonautoregressivespokendialogue}, a corpus of monaural dialogue audio sourced from Internet containing diverse real-world acoustic conditions. We randomly select 100 dialogue sessions for evaluation. 
We did not crop the samples as each sample was shorter than 30 seconds.

\paragraph{Evaluation metrics}
We evaluate the quality of restored and separated outputs using both objective and subjective metrics.

\textit{Common metrics.}
\textbf{DNSMOS}~\cite{dnsmos} and \textbf{NISQA}~\cite{nisqa}: Machine-learning-based speech quality predictors, which predicts the mean opinion score (MOS) on overall perceptual quality of denoised speech.
\textbf{Speaker similarity (Spk.\ Sim.)}: Cosine similarity between speaker embeddings of the output and the reference, computed using the WavLM-based speaker verification model~\cite{wavlm}\footnote{\url{https://hf.co/microsoft/wavlm-base-plus-sv}}.
\textbf{VAD accuracy (VAD Acc.)}: The frame-level voice activity detection (VAD) accuracy on the separated output, used for English and multilingual evaluation, computed using Silero-VAD\footnote{\url{https://github.com/snakers4/silero-vad}}. The reference VAD labels differ by corpus: for SWB, word-level timestamps from the transcription labels are used; for CallFriend, VAD is run on the ground-truth audio to obtain pseudo-labels; 
Note that for OpenDialog, neither speaker-wise tracks nor speaker voice activity labels were provided and we do not report the result.

\textit{Per-dataset intelligibility metrics.}
All intelligibility metrics are computed using Whisper-large-v3 as the ASR model.
\textbf{WER}: Word error rate, used for English (SWB) and in-the-wild data (OpenDialog) evaluation where speaker-wise reference transcripts are available.
\textbf{p-CER}: Pseudo-character error rate, used for the multilingual (CallFriend) evaluation as transcriptions were not available in standard orthography for some languages. Because p-CER is biased by ASR performance on the original recording, it should be interpreted as a relative metric rather than an absolute measure of intelligibility.

\textit{Subjective metric.}
\textbf{MOS}: We conducted a subjective listening test on SWB and OpenDialog. For SWB we compared Noisy, Sidon, GENESES and DialogueSidon and for OpenDialog, we compared GENESES (orig), GENESES and DialogueSidon. Each rater assessed 12 samples. ($360$ ratings per method), scoring recording and separation quality on a 5-point scale. Specifically, each rater was provided with speaker-separated recordings and asked to rate each 20-second sample based on (i) the quality of recording and (ii) speaker consistency in each track. All raters were  recruited through Prolific\footnote{\url{https://www.prolific.com/}} and asked to wear headphones.
The number of raters was 120 for SWB and 90 for OpenDialog.

\paragraph{Baselines}
We compare DialogueSidon against four methods.
\textbf{Noisy} denotes the original telephone recordings without any processing, which serve as an upper bound for speaker identity and VAD structure but contain telephone-band degradations.
\textbf{Sidon}~\cite{nakata2026sidonfastrobustopensource} is a speech restoration model that enhances the monaural mixture as a whole without performing speaker separation, meaning the output is a single restored track rather than speaker-wise signals.
\textbf{GENESES} and \textbf{GENESES (orig)}~\cite{asai2026genesesunifiedgenerativespeech} are unified speech separation and restoration models. GENESES is a flow-matching-based model reported to be robust to complex degradations.
We report results using both their official checkpoint, denoted by (orig), and a retrained model.
For retraining, we trained GENESES from scratch on the same dialogue dataset as DialogueSidon using the hyperparameters reported in the original paper for fair comparison.
At inference time, we used 100 inference steps, as suggested in the paper~\cite{asai2026genesesunifiedgenerativespeech}.

\begin{table*}[!ht]
\caption{
Multilingual evaluation on CallFriend corpus across five languages.\label{tab:multilingual}
We compare Noisy, Sidon, GENESES, and DialogueSidon with 32-dimensional VAE latents.
$^\dagger$Noisy and Sidon do not perform speech separation.
\textbf{Bold}  denote the best and worst among separation methods, respectively.
}
\vspace{-3mm}
\centering
\footnotesize
\scalebox{0.9}{
\begin{tabular}{ll|c|cc|c|c}
\toprule
Language & Method & p-CER (\%) $\downarrow$ & NISQA $\uparrow$ & DNSMOS $\uparrow$ & Spk.\ Sim.\ $\uparrow$ & VAD Acc.\ $\uparrow$ \\
& & \textit{\scriptsize Content} & \multicolumn{2}{c|}{\textit{\scriptsize Recording Quality}} & \textit{\scriptsize Speaker} & \textit{\scriptsize Separation} \\
\midrule
\multirow[t]{4}{*}{German}
 & Noisy$^\dagger$ & --- & 2.131 & 3.001 & --- & --- \\
 & Sidon$^\dagger$ & 11.860 & 3.494 & 3.586 & 0.874 & 0.966 \\
& GENESES (orig) & 76.700 & 3.404 & 3.426 & 0.810 & 0.831  \\
 & GENESES & 52.300 & \textbf{3.646} & \textbf{3.570} & 0.850 & 0.947 \\
 & DialogueSidon & \textbf{14.700} & 3.214 & 3.545 & \textbf{0.900} & \textbf{0.954} \\
\cline{1-7}
\multirow[t]{4}{*}{French}
 & Noisy$^\dagger$ & --- & 2.353 & 3.178 & --- & --- \\
 & Sidon$^\dagger$ & 19.280 & 3.839 & 3.792 & 0.893 & 0.962 \\
  & GENESES (orig) & 86.000 & 3.447 & 3.484 & 0.805 & 0.711 \\
 & GENESES & 61.660 & \textbf{3.753} & 3.591 & 0.881 & 0.923 \\
 & DialogueSidon & \textbf{25.440} & 3.369 & \textbf{3.617} & \textbf{0.905} & \textbf{0.945} \\
\cline{1-7}
\multirow[t]{4}{*}{Japanese}
 & Noisy$^\dagger$ & --- & 2.177 & 2.897 & --- & --- \\
 & Sidon$^\dagger$ & 23.110 & 3.366 & 3.455 & 0.904 & 0.901 \\
  & GENESES (orig) & 120.420 & 3.539 & \textbf{3.466} & 0.754 & 0.510\\
 & GENESES & 89.110 & \textbf{3.728} & 3.460 & 0.824 & 0.785 \\
 & DialogueSidon & \textbf{48.780} & 3.161 & 3.312 & \textbf{0.891} & \textbf{0.796} \\
\cline{1-7}
\multirow[t]{4}{*}{Spanish}
 & Noisy$^\dagger$ & --- & 2.235 & 2.919 & --- & --- \\
 & Sidon$^\dagger$ & 10.670 & 3.556 & 3.615 & 0.877 & 0.962 \\
  & GENESES (orig) & 82.910 & \textbf{3.559} & \textbf{3.360} & 0.777 & 0.686\\
 & GENESES & 50.480 & 3.558 & 3.316 & 0.822 & 0.861 \\
 & DialogueSidon & \textbf{18.720} & 3.142 & 3.313 & \textbf{0.858} & \textbf{0.931} \\
\cline{1-7}
\multirow[t]{4}{*}{Mandarin}
 & Noisy$^\dagger$ & --- & 2.380 & 3.214 & --- & --- \\
 & Sidon$^\dagger$ & 25.240 & 3.682 & 3.701 & 0.944 & 0.951 \\
  & GENESES (orig) & 120.870 & 3.678 & 3.554 & 0.808 & 0.542\\
 & GENESES & 124.000 & \textbf{3.949} & \textbf{3.684} & 0.829 & 0.529 \\
 & DialogueSidon & \textbf{59.960} & 3.436 & 3.526 & \textbf{0.868} & \textbf{0.700} \\
\bottomrule
\end{tabular}
}
\vspace{-3mm}
\end{table*}

\begin{table*}[t]
\caption{
Evaluation on OpenDialog (in-the-wild data).\label{tab:opendialog}
$^\dagger$Noisy and Sidon do not perform speech separation; their metrics are computed on the monaural mixture.
For MOS, $\pm$ indicates standard deviation and all pairwise differences are
statistically significant ($p < 0.05$, two-sided $t$-test).
}
\label{tab:voxconverse}
\vspace{-3mm}
\centering
\footnotesize
\scalebox{0.9}{
\begin{tabular}{l|c|cc|c}
\toprule
Method & WER (\%) $\downarrow$ & NISQA $\uparrow$ & DNSMOS $\uparrow$ & MOS$\uparrow$\\
& \textit{\scriptsize Content} & \multicolumn{2}{c|}{\textit{\scriptsize Recording Quality}} & \multicolumn{1}{c}{\textit{\scriptsize Subjective Quality}}\\
\midrule
Noisy$^\dagger$ & --- & 3.826 & 3.615 & --- \\
Sidon$^\dagger$ & --- & 4.676 & 4.131 & --- \\
\midrule
GENESES (orig) & 74.510 & 3.427 & 3.479 & $2.611 \pm 1.157$ \\
GENESES & 43.790 & \textbf{3.809} & \textbf{3.620} & $3.131 \pm 1.060$\\
DialogueSidon & \textbf{13.860} & 3.568 & 3.598 & $\textbf{3.708} \pm 1.006$\\
\bottomrule
\end{tabular}
}
\vspace{-5mm}
\end{table*}

\subsection{Results}\label{sec:results}

\paragraph{Effect of latent size}
\Cref{tab:latent_size} presents results on the SWB evaluation set for different latent dimensionalities $D$. All DialogueSidon variants substantially outperform the baselines in WER. The best setting, $D = 32$, achieves 14.39\% WER, compared with 33.54\% for the retrained GENESES and 57.47\% for Sidon. Although DialogueSidon does not achieve the best predicted perceptual quality in terms of NISQA or DNSMOS, it better preserves spoken content and speaker characteristics than GENESES, as reflected by WER and speaker similarity. Notably, GENESES (orig)---the original checkpoint trained on monologue mixtures---performs substantially worse than the retrained GENESES across all metrics (e.g., 79.99\% vs.\ 33.54\% WER), confirming that training on conversational data is critical for dialogue separation. The $D = 128$ setting degrades across all metrics, suggesting that excessive latent capacity is detrimental in this setup. Based on these results, we use $D = 32$ in the subsequent experiments.

\paragraph{Subjective evaluation}
To further assess restoration and separation quality, we conducted a human listening test on the SWB evaluation set. As shown in \Cref{tab:mos}, DialogueSidon achieves the highest MOS of 3.895, significantly outperforming GENESES (3.482), Sidon (3.289), and the unprocessed recordings (2.815) ($p < 0.05$, two-sided t-test). These results indicate that, despite not always achieving the best predicted perceptual quality according to NISQA and DNSMOS, DialogueSidon produces speaker-wise outputs that are preferred by human listeners in terms of overall recording quality and speaker consistency.

\paragraph{Multilingual evaluation}
\Cref{tab:multilingual} presents multilingual results on five CallFriend language varieties. DialogueSidon consistently outperforms GENESES in p-CER across all evaluated languages, indicating  better preservation of spoken content. In addition, DialogueSidon generally achieves higher speaker similarity and VAD accuracy, suggesting more faithful speaker-wise reconstruction. Although GENESES often attains higher NISQA or DNSMOS scores, its much worse p-CER indicates a stronger tendency to distort or remove linguistic information. GENESES (orig) performs consistently worse than the retrained GENESES across all languages and metrics, further confirming that the gap between the two reflects the benefit of retraining on conversational dialogue data. These results suggest that DialogueSidon provides a better tradeoff for full-duplex dialogue data construction, where preserving content and speaker structure is critical.

\paragraph{In-the-wild evaluation}
To evaluate DialogueSidon on in-the-wild data, we apply all models to OpenDialog, a corpus of monaural dialogue audio sourced from the Internet. \Cref{tab:opendialog} presents the results. Note that Noisy and Sidon do not produce separated speaker tracks. Therefore WER cannot be measured.
DialogueSidon achieves the lowest WER, substantially outperforming both GENESES variants, consistent with the English and multilingual evaluations. GENESES (orig) shows markedly worse WER than the retrained GENESES (74.51\% vs.\ 43.79\%), again demonstrating the importance of conversational training data for in-the-wild dialogue. However, predicted perceptual quality (NISQA, DNSMOS) is lower for DialogueSidon than GENESES. To further validate sound quality on in-the-wild dialogue, we also report subjective evaluation result (MOS). The results were statistically significant between all compared methods. DialogueSidon achieved the highest MOS, which confirms that DialogueSidon produces higher-quality separation and restoration results than GENESES as judged by human listeners.

\paragraph{Inference efficiency}
For large-scale full-duplex dialogue data construction, inference efficiency is also important. We compared runtime on a single NVIDIA H100 GPU for a 20-second input. DialogueSidon achieves an RTF of 0.010, compared with 0.604 for GENESES, corresponding to a 60.4× speedup. This efficiency advantage likely stems from the relatively small diffusion model used in DialogueSidon (88M parameters), compared with GENESES (393M parameters).


\section{Conclusion}\label{sec:conclusion}
We presented DialogueSidon, a model that jointly restores and separates degraded monaural two-speaker dialogue audio into clean speaker-wise tracks by combining an SSL-VAE latent space with a diffusion-based latent predictor. Experiments on English, multilingual, and in-the-wild dialogue datasets show that DialogueSidon substantially improves content preservation and separation quality over a baseline with 60$\times$ faster inference, suggesting a practical path toward constructing full-duplex dialogue resources from large-scale in-the-wild audio. Future work includes extending to more speakers and broader language coverage.

\ifcameraready
\section*{Acknowledgments}
This work was supported by JST Moonshot JPMJMS2011, JST BOOST JPMJBY24C9, JSPS KAKENHI, Grant Number 25KJ0806, and the AIST policy-based budget project ``R\&D on Generative AI Foundation Models for the Physical Domain.''
\fi

\bibliography{custom}

@INPROCEEDINGS{wang23tf-gridnet,
  author={Wang, Zhong-Qiu and Cornell, Samuele and Choi, Shukjae and Lee, Younglo and Kim, Byeong-Yeol and Watanabe, Shinji},
  booktitle={ICASSP 2023 - 2023 IEEE International Conference on Acoustics, Speech and Signal Processing (ICASSP)}, 
  title={{TF-GRIDNET}: Making Time-Frequency Domain Models Great Again for Monaural Speaker Separation}, 
  year={2023},
  volume={},
  number={},
  pages={1-5},
  doi={10.1109/ICASSP49357.2023.10094992}}

@INPROCEEDINGS{miipher,
  author={Koizumi, Yuma and Zen, Heiga and Karita, Shigeki and Ding, Yifan and Yatabe, Kohei and Morioka, Nobuyuki and Zhang, Yu and Han, Wei and Bapna, Ankur and Bacchiani, Michiel},
  booktitle={2023 IEEE Workshop on Applications of Signal Processing to Audio and Acoustics (WASPAA)}, 
  title={Miipher: A Robust Speech Restoration Model Integrating Self-Supervised Speech and Text Representations}, 
  year={2023},
  volume={},
  number={},
  pages={1-5},
  doi={10.1109/WASPAA58266.2023.10248089}}

@INPROCEEDINGS{karita2025miipher,
  author={Karita, Shigeki and Koizumi, Yuma and Zen, Heiga and Ishikawa, Haruko and Scheibler, Robin and Bacchiani, Michiel},
  booktitle={2025 IEEE Workshop on Applications of Signal Processing to Audio and Acoustics (WASPAA)}, 
  title={Miipher-2: A Universal Speech Restoration Model for Million-Hour Scale Data Restoration}, 
  year={2025},
  volume={},
  number={},
  pages={1-5},
  doi={10.1109/WASPAA66052.2025.11230923}}

@INPROCEEDINGS{dnsmos,
  author={Reddy, Chandan K A and Gopal, Vishak and Cutler, Ross},
  booktitle={ICASSP 2021 - 2021 IEEE International Conference on Acoustics, Speech and Signal Processing (ICASSP)}, 
  title={Dnsmos: A Non-Intrusive Perceptual Objective Speech Quality Metric to Evaluate Noise Suppressors}, 
  year={2021},
  volume={},
  number={},
  pages={6493-6497},
  doi={10.1109/ICASSP39728.2021.9414878}}

@inproceedings{nisqa,
  title     = {{NISQA: A Deep CNN-Self-Attention Model for Multidimensional Speech Quality Prediction with Crowdsourced Datasets}},
  author    = {Gabriel Mittag and Babak Naderi and Assmaa Chehadi and Sebastian Möller},
  year      = {2021},
  booktitle = {{Interspeech 2021}},
  pages     = {2127--2131},
  doi       = {10.21437/Interspeech.2021-299},
  issn      = {2958-1796},
}

@inproceedings{Wichern2019WHAM,
    title     = {{WHAM!}: Extending Speech Separation to Noisy Environments},
    author    = {Wichern, Gordon and Antognini, Joe and others },
    booktitle = {Proc. Interspeech},
    year      = {2019},
}

@ARTICLE{fsd50k,
  author={Fonseca, Eduardo and Favory, Xavier and Pons, Jordi and Font, Frederic and Serra, Xavier},
  journal={IEEE/ACM Transactions on Audio, Speech, and Language Processing}, 
  title={{FSD50K}: An Open Dataset of Human-Labeled Sound Events}, 
  year={2022},
  volume={30},
  number={},
  pages={829-852},
  doi={10.1109/TASLP.2021.3133208}}

@INPROCEEDINGS{windnoise,
  author={Mirabilii, Daniele and Lodermeyer, Alexander and Czwielong, Felix and Becker, Stefan and Habets, Emanuël A.P.},
  booktitle={2022 International Workshop on Acoustic Signal Enhancement (IWAENC)}, 
  title={Simulating Wind Noise with Airflow Speed-Dependent Characteristics}, 
  year={2022},
  volume={},
  number={},
  pages={1-5},
  doi={10.1109/IWAENC53105.2022.9914785}}

@ARTICLE{wavlm,
  author={Chen, Sanyuan and Wang, Chengyi and Chen, Zhengyang and Wu, Yu and Liu, Shujie and Chen, Zhuo and Li, Jinyu and Kanda, Naoyuki and Yoshioka, Takuya and Xiao, Xiong and Wu, Jian and Zhou, Long and Ren, Shuo and Qian, Yanmin and Qian, Yao and Wu, Jian and Zeng, Michael and Yu, Xiangzhan and Wei, Furu},
  journal={IEEE Journal of Selected Topics in Signal Processing}, 
  title={{WavLM}: Large-Scale Self-Supervised Pre-Training for Full Stack Speech Processing}, 
  year={2022},
  volume={16},
  number={6},
  pages={1505--1518},
}

@inproceedings{
loshchilov2018decoupled,
title={Decoupled Weight Decay Regularization},
author={Ilya Loshchilov and Frank Hutter},
booktitle={International Conference on Learning Representations},
year={2019},
url={https://openreview.net/forum?id=Bkg6RiCqY7},
}

@misc{whisper,
      title={Robust Speech Recognition via Large-Scale Weak Supervision}, 
      author={Alec Radford and Jong Wook Kim and Tao Xu and Greg Brockman and Christine McLeavey and Ilya Sutskever},
      year={2022},
      eprint={2212.04356},
      archivePrefix={arXiv},
      primaryClass={eess.AS},
      url={https://arxiv.org/abs/2212.04356}, 
}

@inproceedings{koizumi23_interspeech,
  title     = {{LibriTTS-R: A Restored Multi-Speaker Text-to-Speech Corpus}},
  author    = {Yuma Koizumi and Heiga Zen and Shigeki Karita and Yifan Ding and Kohei Yatabe and Nobuyuki Morioka and Michiel Bacchiani and Yu Zhang and Wei Han and Ankur Bapna},
  year      = {2023},
  booktitle = {{Interspeech 2023}},
  pages     = {5496--5500},
  doi       = {10.21437/Interspeech.2023-1584},
  issn      = {2958-1796},
}

@inproceedings{ma24c_interspeech,
  title     = {{FLEURS-R: A Restored Multilingual Speech Corpus for Generation Tasks}},
  author    = {Min Ma and Yuma Koizumi and Shigeki Karita and Heiga Zen and Jason Riesa and Haruko Ishikawa and Michiel Bacchiani},
  year      = {2024},
  booktitle = {{Interspeech 2024}},
  pages     = {1835--1839},
  doi       = {10.21437/Interspeech.2024-1356},
  issn      = {2958-1796},
}

@misc{Guimares2025DiTSEHG,
      title={{DiTSE}: High-Fidelity Generative Speech Enhancement via Latent Diffusion Transformers}, 
      author={Heitor R. Guimarães and Jiaqi Su and Rithesh Kumar and Tiago H. Falk and Zeyu Jin},
      year={2025},
      eprint={2504.09381},
      archivePrefix={arXiv},
      primaryClass={eess.AS},
      url={https://arxiv.org/abs/2504.09381}, 
}

@inproceedings{Kong2020HiFi-GAN:Synthesis,
 author = {Kong, Jungil and Kim, Jaehyeon and Bae, Jaekyoung},
 booktitle = {Advances in Neural Information Processing Systems},
 editor = {H. Larochelle and M. Ranzato and R. Hadsell and M.F. Balcan and H. Lin},
 pages = {17022--17033},
 publisher = {Curran Associates, Inc.},
 title = {{HiFi-GAN}: Generative Adversarial Networks for Efficient and High Fidelity Speech Synthesis},
 url = {https://proceedings.neurips.cc/paper_files/paper/2020/file/c5d736809766d46260d816d8dbc9eb44-Paper.pdf},
 volume = {33},
 year = {2020}

}

@inproceedings{gemmeke2017audio,
  author={Gemmeke, Jort F. and Ellis, Daniel P. W. and Freedman, Dylan and Jansen, Aren and Lawrence, Wade and Moore, R. Channing and Plakal, Manoj and Ritter, Marvin},
  booktitle={2017 IEEE International Conference on Acoustics, Speech and Signal Processing (ICASSP)}, 
  title={Audio Set: An ontology and human-labeled dataset for audio events}, 
  year={2017},
  volume={},
  number={},
  pages={776-780},
  doi={10.1109/ICASSP.2017.7952261}}

@inproceedings{scheibler2018pyroomacoustics,
  author={Scheibler, Robin and Bezzam, Eric and Dokmanić, Ivan},
  booktitle={2018 IEEE International Conference on Acoustics, Speech and Signal Processing (ICASSP)}, 
  title={Pyroomacoustics: A Python Package for Audio Room Simulation and Array Processing Algorithms}, 
  year={2018},
  volume={},
  number={},
  pages={351-355},
  doi={10.1109/ICASSP.2018.8461310}}

@inproceedings{liu22y_interspeech,
  title     = {{VoiceFixer: A Unified Framework for High-Fidelity Speech Restoration}},
  author    = {Haohe Liu and Xubo Liu and Qiuqiang Kong and Qiao Tian and Yan Zhao and DeLiang Wang and Chuanzeng Huang and Yuxuan Wang},
  year      = {2022},
  booktitle = {{Interspeech 2022}},
  pages     = {4232--4236},
  doi       = {10.21437/Interspeech.2022-11026},
  issn      = {2958-1796},
}

@INPROCEEDINGS{hiresldm,
  author={Dhyani, Tushar and Lux, Florian and Mancusi, Michele and Fabbro, Giorgio and Hohl, Fritz and Vu, Ngoc Thang},
  booktitle={ICASSP 2025 - 2025 IEEE International Conference on Acoustics, Speech and Signal Processing (ICASSP)}, 
  title={High-Resolution Speech Restoration with Latent Diffusion Model}, 
  year={2025},
  volume={},
  number={},
  pages={1-5},
  doi={10.1109/ICASSP49660.2025.10890277}}

@misc{Communication2023SeamlessME,
      title={Seamless: Multilingual Expressive and Streaming Speech Translation}, 
      author={Seamless Communication and Loïc Barrault and Yu-An Chung and Mariano Coria Meglioli and David Dale and Ning Dong and Mark Duppenthaler and Paul-Ambroise Duquenne and Brian Ellis and Hady Elsahar and Justin Haaheim and John Hoffman and Min-Jae Hwang and Hirofumi Inaguma and Christopher Klaiber and Ilia Kulikov and Pengwei Li and Daniel Licht and Jean Maillard and Ruslan Mavlyutov and Alice Rakotoarison and Kaushik Ram Sadagopan and Abinesh Ramakrishnan and Tuan Tran and Guillaume Wenzek and Yilin Yang and Ethan Ye and Ivan Evtimov and Pierre Fernandez and Cynthia Gao and Prangthip Hansanti and Elahe Kalbassi and Amanda Kallet and Artyom Kozhevnikov and Gabriel Mejia Gonzalez and Robin San Roman and Christophe Touret and Corinne Wong and Carleigh Wood and Bokai Yu and Pierre Andrews and Can Balioglu and Peng-Jen Chen and Marta R. Costa-jussà and Maha Elbayad and Hongyu Gong and Francisco Guzmán and Kevin Heffernan and Somya Jain and Justine Kao and Ann Lee and Xutai Ma and Alex Mourachko and Benjamin Peloquin and Juan Pino and Sravya Popuri and Christophe Ropers and Safiyyah Saleem and Holger Schwenk and Anna Sun and Paden Tomasello and Changhan Wang and Jeff Wang and Skyler Wang and Mary Williamson},
      year={2023},
      eprint={2312.05187},
      archivePrefix={arXiv},
      primaryClass={cs.CL},
      url={https://arxiv.org/abs/2312.05187}, 
}

@inproceedings{
hu2022lora,
title={Lo{RA}: Low-Rank Adaptation of Large Language Models},
author={Edward J Hu and yelong shen and Phillip Wallis and Zeyuan Allen-Zhu and Yuanzhi Li and Shean Wang and Lu Wang and Weizhu Chen},
booktitle={International Conference on Learning Representations},
year={2022},
url={https://openreview.net/forum?id=nZeVKeeFYf9}
}

@inproceedings{
kumar2023highfidelity,
title={High-Fidelity Audio Compression with Improved {RVQGAN}},
author={Rithesh Kumar and Prem Seetharaman and Alejandro Luebs and Ishaan Kumar and Kundan Kumar},
booktitle={Thirty-seventh Conference on Neural Information Processing Systems},
year={2023},
url={https://openreview.net/forum?id=qjnl1QUnFA}
}

@inproceedings{saijo25_interspeech,
  title     = {{Interspeech 2025 URGENT Speech Enhancement Challenge}},
  author    = {Kohei Saijo and Wangyou Zhang and Samuele Cornell and Robin Scheibler and Chenda Li and Zhaoheng Ni and Anurag Kumar and Marvin Sach and Yihui Fu and Wei Wang and Tim Fingscheidt and Shinji Watanabe},
  year      = {2025},
  booktitle = {{Interspeech 2025}},
  pages     = {858--862},
  doi       = {10.21437/Interspeech.2025-1363},
  issn      = {2958-1796},
}

@article{imagesource,
    author = {Allen, Jont B. and Berkley, David A.},
    title = {Image method for efficiently simulating small‐room acoustics},
    journal = {The Journal of the Acoustical Society of America},
    volume = {65},
    number = {4},
    pages = {943-950},
    year = {1979},
    month = {04},
    issn = {0001-4966},
    doi = {10.1121/1.382599},
    url = {https://doi.org/10.1121/1.382599},
    eprint = {https://pubs.aip.org/asa/jasa/article-pdf/65/4/943/11426543/943_1_online.pdf},

}

@article{usm,
  journal={arXiv},
  author={Yu Zhang and Wei Han and others},
  title={Google {USM}: Scaling Automatic Speech Recognition Beyond 100 Languages},
  year={2023},
  volume={abs/2303.01037},
}

@misc{nakata2026jchatjapaneselargescalespoken,
      title={{J-CHAT}: Japanese Large-scale Spoken Dialogue Corpus for Spoken Dialogue Language Modeling}, 
      author={Wataru Nakata and Kentaro Seki and Hitomi Yanaka and Yuki Saito and Shinnosuke Takamichi and Hiroshi Saruwatari},
      year={2026},
      eprint={2407.15828},
      archivePrefix={arXiv},
      primaryClass={cs.CL},
      url={https://arxiv.org/abs/2407.15828}, 
}

@misc{defossez2024moshispeechtextfoundationmodel,
      title={Moshi: a speech-text foundation model for real-time dialogue}, 
      author={Alexandre Défossez and Laurent Mazaré and Manu Orsini and Amélie Royer and Patrick Pérez and Hervé Jégou and Edouard Grave and Neil Zeghidour},
      year={2024},
      eprint={2410.00037},
      archivePrefix={arXiv},
      primaryClass={eess.AS},
      url={https://arxiv.org/abs/2410.00037}, 
}

@article{nguyen-etal-2023-generative,
    title = "Generative Spoken Dialogue Language Modeling",
    author = "Nguyen, Tu Anh  and
      Kharitonov, Eugene  and
      Copet, Jade  and
      Adi, Yossi  and
      Hsu, Wei-Ning  and
      Elkahky, Ali  and
      Tomasello, Paden  and
      Algayres, Robin  and
      Sagot, Beno{\^i}t  and
      Mohamed, Abdelrahman  and
      Dupoux, Emmanuel",
    journal = "Transactions of the Association for Computational Linguistics",
    volume = "11",
    year = "2023",
    address = "Cambridge, MA",
    publisher = "MIT Press",
    url = "https://aclanthology.org/2023.tacl-1.15/",
    doi = "10.1162/tacl_a_00545",
    pages = "250--266"
}

@misc{roy2026personaplexvoicerolecontrol,
      title={PersonaPlex: Voice and Role Control for Full Duplex Conversational Speech Models}, 
      author={Rajarshi Roy and Jonathan Raiman and Sang-gil Lee and Teodor-Dumitru Ene and Robert Kirby and Sungwon Kim and Jaehyeon Kim and Bryan Catanzaro},
      year={2026},
      eprint={2602.06053},
      archivePrefix={arXiv},
      primaryClass={cs.CL},
      url={https://arxiv.org/abs/2602.06053}, 
}

@inproceedings{cieri-etal-2004-fisher,
    title = "The Fisher Corpus: a Resource for the Next Generations of Speech-to-Text",
    author = "Cieri, Christopher  and
      Miller, David  and
      Walker, Kevin",
    editor = "Lino, Maria Teresa  and
      Xavier, Maria Francisca  and
      Ferreira, F{\'a}tima  and
      Costa, Rute  and
      Silva, Raquel",
    booktitle = "Proceedings of the Fourth International Conference on Language Resources and Evaluation ({LREC}{'}04)",
    month = may,
    year = "2004",
    address = "Lisbon, Portugal",
    publisher = "European Language Resources Association (ELRA)",
    url = "https://aclanthology.org/L04-1500/"
}

@misc{ldc2008_cabank_callhome_japanese,
  author       = {{LDC}},
  title        = {{CABank Japanese CallHome Corpus}},
  year         = {2008},
  publisher    = {TalkBank},
  doi          = {10.21415/T5H59V},
  url          = {https://doi.org/10.21415/T5H59V},
}

@misc{ldc2008_cabank_callhome_english,
  author       = {{LDC}},
  title        = {{CABank English CallHome Corpus}},
  year         = {2008},
  publisher    = {TalkBank},
  doi          = {10.21415/T5KP54},
  url          = {https://doi.org/10.21415/T5KP54},
}

@misc{ldc2008_cabank_callhome_chinese,
  author       = {{LDC}},
  title        = {{CABank Chinese CallHome Corpus}},
  year         = {2008},
  publisher    = {TalkBank},
  doi          = {10.21415/T54022},
  url          = {https://doi.org/10.21415/T54022},
}

@misc{ldc2008_cabank_callhome_spanish,
  author       = {{LDC}},
  title        = {{CABank Spanish CallHome Corpus}},
  year         = {2008},
  publisher    = {TalkBank},
  doi          = {10.21415/T51K54},
  url          = {https://doi.org/10.21415/T51K54},
}

@misc{godfrey1993_switchboard1_release2,
  author       = {Godfrey, John J. and Holliman, Edward},
  title        = {{Switchboard-1 Release 2}},
  year         = {1993},
  publisher    = {LDC},
  address      = {Philadelphia, PA, USA},
  doi          = {10.35111/sw3h-rw02},
  url          = {https://doi.org/10.35111/sw3h-rw02}
}

@misc{seamless_interaction,
      title={Seamless Interaction: Dyadic Audiovisual Motion Modeling and Large-Scale Dataset}, 
      author={Vasu Agrawal and Akinniyi Akinyemi and Kathryn Alvero and Morteza Behrooz and Julia Buffalini and Fabio Maria Carlucci and Joy Chen and Junming Chen and Zhang Chen and Shiyang Cheng and Praveen Chowdary and Joe Chuang and Antony D'Avirro and Jon Daly and Ning Dong and Mark Duppenthaler and Cynthia Gao and Jeff Girard and Martin Gleize and Sahir Gomez and Hongyu Gong and Srivathsan Govindarajan and Brandon Han and Sen He and Denise Hernandez and Yordan Hristov and Rongjie Huang and Hirofumi Inaguma and Somya Jain and Raj Janardhan and Qingyao Jia and Christopher Klaiber and Dejan Kovachev and Moneish Kumar and Hang Li and Yilei Li and Pavel Litvin and Wei Liu and Guangyao Ma and Jing Ma and Martin Ma and Xutai Ma and Lucas Mantovani and Sagar Miglani and Sreyas Mohan and Louis-Philippe Morency and Evonne Ng and Kam-Woh Ng and Tu Anh Nguyen and Amia Oberai and Benjamin Peloquin and Juan Pino and Jovan Popovic and Omid Poursaeed and Fabian Prada and Alice Rakotoarison and Rakesh Ranjan and Alexander Richard and Christophe Ropers and Safiyyah Saleem and Vasu Sharma and Alex Shcherbyna and Jia Shen and Jie Shen and Anastasis Stathopoulos and Anna Sun and Paden Tomasello and Tuan Tran and Arina Turkatenko and Bo Wan and Chao Wang and Jeff Wang and Mary Williamson and Carleigh Wood and Tao Xiang and Yilin Yang and Julien Yao and Chen Zhang and Jiemin Zhang and Xinyue Zhang and Jason Zheng and Pavlo Zhyzheria and Jan Zikes and Michael Zollhoefer},
      year={2025},
      eprint={2506.22554},
      archivePrefix={arXiv},
      primaryClass={cs.CV},
      url={https://arxiv.org/abs/2506.22554}, 
}

@misc{canavan1996_callfriend_canadian_french,
  author       = {Canavan, Alexandra and Zipperlen, George},
  title        = {{CALLFRIEND Canadian French}},
  year         = {1996},
  publisher    = {LDC},
  address      = {Philadelphia, PA, USA},
  doi          = {10.35111/91pj-x181},
  url          = {https://doi.org/10.35111/91pj-x181}
}

@misc{canavan1996_callfriend_german,
  author       = {Canavan, Alexandra and Zipperlen, George},
  title        = {{CALLFRIEND German}},
  year         = {1996},
  publisher    = {LDC},
  address      = {Philadelphia, PA, USA},
  doi          = {10.35111/99vs-vv60},
  url          = {https://doi.org/10.35111/99vs-vv60}
}

@misc{canavan1996_callfriend_japanese,
  author       = {Canavan, Alexandra and Zipperlen, George},
  title        = {{CALLFRIEND Japanese}},
  year         = {1996},
  publisher    = {LDC},
  address      = {Philadelphia, PA, USA},
  doi          = {10.35111/wv05-we23},
  url          = {https://doi.org/10.35111/wv05-we23}
}

@misc{canavan1996_callfriend_spanish_non_caribbean,
  author       = {Canavan, Alexandra and Zipperlen, George},
  title        = {{CALLFRIEND Spanish Non-Caribbean Dialect}},
  year         = {1996},
  publisher    = {LDC},
  address      = {Philadelphia, PA, USA},
  doi          = {10.35111/nce5-9n94},
  url          = {https://doi.org/10.35111/nce5-9n94}
}

@misc{tracey2025_bolt_callfriend_callhome_mandarin,
  author       = {Tracey, Jennifer and Graff, David and Chen, Song and Strassel, Stephanie},
  title        = {{BOLT CTS CALLFRIEND CALLHOME Mainland Mandarin Chinese Audio}},
  year         = {2025},
  publisher    = {LDC},
  address      = {Philadelphia, PA, USA},
  doi          = {10.35111/d6y0-25s04},
  url          = {https://doi.org/10.35111/d6y0-25s04}
}

@ARTICLE{convtasnet,
  author={Luo, Yi and Mesgarani, Nima},
  journal={IEEE/ACM Transactions on Audio, Speech, and Language Processing}, 
  title={{Conv-TasNet}: Surpassing Ideal Time–Frequency Magnitude Masking for Speech Separation}, 
  year={2019},
  volume={27},
  number={8},
  pages={1256-1266},
  doi={10.1109/TASLP.2019.2915167}}

@misc{nakata2026sidonfastrobustopensource,
      title={Sidon: Fast and Robust Open-Source Multilingual Speech Restoration for Large-scale Dataset Cleansing}, 
      author={Wataru Nakata and Yuki Saito and Yota Ueda and Hiroshi Saruwatari},
      year={2026},
      eprint={2509.17052},
      archivePrefix={arXiv},
      primaryClass={cs.SD},
      url={https://arxiv.org/abs/2509.17052}, 
}

@misc{asai2026genesesunifiedgenerativespeech,
      title={Geneses: Unified Generative Speech Enhancement and Separation}, 
      author={Kohei Asai and Wataru Nakata and Yuki Saito and Hiroshi Saruwatari},
      year={2026},
      eprint={2601.18456},
      archivePrefix={arXiv},
      primaryClass={cs.SD},
      url={https://arxiv.org/abs/2601.18456}, 
}

@INPROCEEDINGS{noisySS,
  author={Zhang, Zizheng and Chen, Chen and Chen, Hsin-Hung and Liu, Xiang and Hu, Yuchen and Chng, Eng Siong},
  booktitle={ICASSP 2024 - 2024 IEEE International Conference on Acoustics, Speech and Signal Processing (ICASSP)}, 
  title={Noise-Aware Speech Separation with Contrastive Learning}, 
  year={2024},
  volume={},
  number={},
  pages={1381-1385},
  doi={10.1109/ICASSP48485.2024.10448214}}

@inproceedings{peng25c_interspeech,
  title     = {{OWSM v4: Improving Open Whisper-Style Speech Models via Data Scaling and Cleaning}},
  author    = {Yifan Peng and Muhammad Shakeel and Yui Sudo and William Chen and Jinchuan Tian and Chyi-Jiunn Lin and Shinji Watanabe},
  year      = {2025},
  booktitle = {{Interspeech 2025}},
  pages     = {2225--2229},
  doi       = {10.21437/Interspeech.2025-1062},
  issn      = {2958-1796},
}

@INPROCEEDINGS{floss,
  author={Scheibler, Robin and Hershey, John R. and Doucet, Arnaud and Li, Henry},
  booktitle={2025 IEEE Workshop on Applications of Signal Processing to Audio and Acoustics (WASPAA)}, 
  title={Source Separation by Flow Matching}, 
  year={2025},
  volume={},
  number={},
  pages={1-5},
  doi={10.1109/WASPAA66052.2025.11230963}}

@inproceedings{veluri-etal-2024-beyond,
    title = "Beyond Turn-Based Interfaces: Synchronous {LLM}s as Full-Duplex Dialogue Agents",
    author = "Veluri, Bandhav  and
      Peloquin, Benjamin N  and
      Yu, Bokai  and
      Gong, Hongyu  and
      Gollakota, Shyamnath",
    editor = "Al-Onaizan, Yaser  and
      Bansal, Mohit  and
      Chen, Yun-Nung",
    booktitle = "Proceedings of the 2024 Conference on Empirical Methods in Natural Language Processing",
    month = nov,
    year = "2024",
    address = "Miami, Florida, USA",
    publisher = "Association for Computational Linguistics",
    url = "https://aclanthology.org/2024.emnlp-main.1192/",
    doi = "10.18653/v1/2024.emnlp-main.1192",
    pages = "21390--21402"
}

@inproceedings{snake,
 author = {Ziyin, Liu and Hartwig, Tilman and Ueda, Masahito},
 booktitle = {Advances in Neural Information Processing Systems},
 editor = {H. Larochelle and M. Ranzato and R. Hadsell and M.F. Balcan and H. Lin},
 pages = {1583--1594},
 publisher = {Curran Associates, Inc.},
 title = {Neural Networks Fail to Learn Periodic Functions and How to Fix It},
 url = {https://proceedings.neurips.cc/paper_files/paper/2020/file/1160453108d3e537255e9f7b931f4e90-Paper.pdf},
 volume = {33},
 year = {2020}
}

@misc{zhu2025zipvoicedialognonautoregressivespokendialogue,
      title={ZipVoice-Dialog: Non-Autoregressive Spoken Dialogue Generation with Flow Matching}, 
      author={Han Zhu and Wei Kang and Liyong Guo and Zengwei Yao and Fangjun Kuang and Weiji Zhuang and Zhaoqing Li and Zhifeng Han and Dong Zhang and Xin Zhang and Xingchen Song and Long Lin and Daniel Povey},
      year={2025},
      eprint={2507.09318},
      archivePrefix={arXiv},
      primaryClass={eess.AS},
      url={https://arxiv.org/abs/2507.09318}, 
}

@InProceedings{Peebles_2023_ICCV,
  author={Peebles, William and Xie, Saining},
  booktitle={2023 IEEE/CVF International Conference on Computer Vision (ICCV)}, 
  title={Scalable Diffusion Models with Transformers}, 
  year={2023},
  volume={},
  number={},
  pages={4172-4182},
  doi={10.1109/ICCV51070.2023.00387}}

@article{Lu_2025,
   title={{DPM-Solver++}: Fast Solver for Guided Sampling of Diffusion Probabilistic Models},
   volume={22},
   ISSN={2731-5398},
   url={http://dx.doi.org/10.1007/s11633-025-1562-4},
   DOI={10.1007/s11633-025-1562-4},
   number={4},
   journal={Machine Intelligence Research},
   publisher={Springer Science and Business Media LLC},
   author={Lu, Cheng and Zhou, Yuhao and Bao, Fan and Chen, Jianfei and Li, Chongxuan and Zhu, Jun},
   year={2025},
   month=jun, pages={730–751} }

@inproceedings{
salimans2022progressive,
title={Progressive Distillation for Fast Sampling of Diffusion Models},
author={Tim Salimans and Jonathan Ho},
booktitle={International Conference on Learning Representations},
year={2022},
url={https://openreview.net/forum?id=TIdIXIpzhoI}
}

@INPROCEEDINGS{pit,
  author={Yu, Dong and Kolbæk, Morten and Tan, Zheng-Hua and Jensen, Jesper},
  booktitle={2017 IEEE International Conference on Acoustics, Speech and Signal Processing (ICASSP)}, 
  title={Permutation invariant training of deep models for speaker-independent multi-talker speech separation}, 
  year={2017},
  volume={},
  number={},
  pages={241-245},
  doi={10.1109/ICASSP.2017.7952154}}

@misc{shi2025samaudiosegmentaudio,
      title={{SAM Audio}: Segment Anything in Audio}, 
      author={Bowen Shi and Andros Tjandra and John Hoffman and Helin Wang and Yi-Chiao Wu and Luya Gao and Julius Richter and Matt Le and Apoorv Vyas and Sanyuan Chen and Christoph Feichtenhofer and Piotr Dollár and Wei-Ning Hsu and Ann Lee},
      year={2025},
      eprint={2512.18099},
      archivePrefix={arXiv},
      primaryClass={eess.AS},
      url={https://arxiv.org/abs/2512.18099}, 
}

@INPROCEEDINGS{dailytalk,
  author={Lee, Keon and Park, Kyumin and Kim, Daeyoung},
  booktitle={ICASSP 2023 - 2023 IEEE International Conference on Acoustics, Speech and Signal Processing (ICASSP)}, 
  title={DailyTalk: Spoken Dialogue Dataset for Conversational Text-to-Speech}, 
  year={2023},
  volume={},
  number={},
  pages={1-5},
  doi={10.1109/ICASSP49357.2023.10095751}}

@INPROCEEDINGS{emilia,
  author={He, Haorui and Shang, Zengqiang and Wang, Chaoren and Li, Xuyuan and Gu, Yicheng and Hua, Hua and Liu, Liwei and Yang, Chen and Li, Jiaqi and Shi, Peiyang and Wang, Yuancheng and Chen, Kai and Zhang, Pengyuan and Wu, Zhizheng},
  booktitle={2024 IEEE Spoken Language Technology Workshop (SLT)}, 
  title={Emilia: An Extensive, Multilingual, and Diverse Speech Dataset For Large-Scale Speech Generation}, 
  year={2024},
  volume={},
  number={},
  pages={885-890},
  doi={10.1109/SLT61566.2024.10832365}}

@article{rope,
title = {RoFormer: Enhanced transformer with Rotary Position Embedding},
journal = {Neurocomputing},
volume = {568},
pages = {127063},
year = {2024},
issn = {0925-2312},
doi = {https://doi.org/10.1016/j.neucom.2023.127063},
url = {https://www.sciencedirect.com/science/article/pii/S0925231223011864},
author = {Jianlin Su and Murtadha Ahmed and Yu Lu and Shengfeng Pan and Wen Bo and Yunfeng Liu}
}

\appendix
\section{Details of the degradation pipeline}\label{sec:deg}
We used the degradation pipeline in the following order.
   \begin{enumerate}

        \item Reverberation: We used pyroomacoustics~\cite{scheibler2018pyroomacoustics}
            for simulating room impulse responses (RIRs). Specifically, random RT60 and rectangular cuboid room dimensions were drawn from $\mathcal{U}(0.1,1.0)$ seconds and $\mathcal{U}(2,20)$~m respectively. Based on the drawn RT60 and room dimensions, wall absorption and maximum order of the image-source method~\cite{imagesource} were calculated using Sabine's equation. Then, RIRs were simulated.

\item Background noise: We formed a noise pool from AudioSet~\cite{gemmeke2017audio},
Free Music Archive, WHAM!~\cite{Wichern2019WHAM}, FSD50K~\cite{fsd50k}, and synthetic
wind noise generated by SC-Wind-Noise-Generator~\cite{windnoise}. 
For each clean utterance, we randomly sampled a single noise recording from this pool.
The selected noise was looped to exceed the utterance duration and then truncated to match exactly, and it was added at an SNR drawn from $\mathcal{U}(-5,20)$~dB.

        \item Band limitation: The input speech was randomly resampled at \{8,
            16, 22.05, 24, 44.1, 48\}~kHz sampling rate before being converted
            back to the original sampling rate.

        \item Clipping: The input speech was randomly clipped by setting its new
            minimum value to the value corresponding to a quantile uniformly
            chosen between the 0th and 10th percentiles, and its new maximum value
            to the value corresponding to a quantile uniformly chosen between the
            90th and 100th percentiles of the original signal.

        \item Codec: We applied the MP3 compression with a random average
            bitrate ranging from 65~kbps to 245~kbps.

        \item Packet loss: Random 9\% segments of speech were selected for packet
            loss. For each segment duration sampled from $\mathcal{U}(20,200)$~milliseconds
            were selected to be replaced with zeros to simulate packet loss.
        \item Mixing: From the degraded tracks $\tilde{y}_1, \tilde{y}_2$, monaural dialogue audio was created as a weighted sum $x = w \cdot \tilde{y}_1 + (1 - w) \cdot \tilde{y}_2$, where the weight $w$ is drawn from $\mathcal{U}(0.3,0.7)$.
\end{enumerate}

\end{document}